\documentclass[prodmode,acmcsur]{acmsmall} 

\usepackage{amsmath}%
\usepackage{amsfonts}%
\usepackage{amssymb}%
\usepackage{graphicx}
\usepackage{array}

\usepackage[flushleft]{threeparttable}
\usepackage[hidelinks]{hyperref}

\DeclareMathOperator*{\argmax}{arg\,max}
\DeclareMathOperator*{\argmin}{arg\,min}

\usepackage[ruled]{algorithm2e}

\SetAlFnt{\small}
\SetAlCapFnt{\small}
\SetAlCapNameFnt{\small}
\SetAlCapHSkip{0pt}
\IncMargin{-\parindent}

\acmVolume{X}
\acmNumber{Y}
\acmArticle{Z}
\acmYear{2019}
\acmMonth{5}

\setcopyright{rightsretained}

\doi{0000001.0000001}

\issn{1234-56789}

\begin{document}

\markboth{S. Fletcher \& M.Z. Islam}{Decision Tree Classification with Differential Privacy: A Survey}

\title{Decision Tree Classification with Differential Privacy: A Survey}
\author{SAM FLETCHER
\affil{Charles Sturt University}
MD ZAHIDUL ISLAM
\affil{Charles Sturt University}}

\begin{abstract}
Data mining information about people is becoming increasingly important in the data-driven society of the 21st century. Unfortunately, sometimes there are real-world considerations that conflict with the goals of data mining; sometimes the privacy of the people being data mined needs to be considered. This necessitates that the output of data mining algorithms be modified to preserve privacy while simultaneously not ruining the predictive power of the outputted model. Differential privacy is a strong, enforceable definition of privacy that can be used in data mining algorithms, guaranteeing that nothing will be learned about the people in the data that could not already be discovered without their participation. In this survey, we focus on one particular data mining algorithm -- decision trees -- and how differential privacy interacts with each of the components that constitute decision tree algorithms. We analyze both greedy and random decision trees, and the conflicts that arise when trying to balance privacy requirements with the accuracy of the model.
\end{abstract}

%
%

%



\acmformat{Sam Fletcher and Md Zahidul Islam, 2019. Decision tree classification with differential privacy: a survey.}

\begin{bottomstuff}

Authors' email: sam.pt.fletcher@gmail.com and zislam@csu.edu.au
\end{bottomstuff}

\maketitle

\section{Introduction}

Gathering and analyzing information about people is becoming increasingly important in the data-driven society of the 21st century. Technology continues to facilitate new and efficient ways of collecting data, but extracting knowledge from the data remains a difficult and nuanced topic. Many fields of science intersect when ``mining'' data for information; machine learning, statistics and database systems all play a role in producing useful information from (potentially enormous) repositories of data \cite{Breiman2001a}. The intersection of these fields is often referred to as ``data mining'', and data mining algorithms cover a wide range of applications; some output humanly-understandable patterns discovered in the data \cite{Vellido2012}, some produce a classification or regression model that can make predictions about the future \cite{Criminisi2011}, and others detect anomalies in the data \cite{Chandola2009}. In this survey, we focus on one particular kind of classification model: decision trees \cite{Quinlan1986}. We introduce decision trees in detail in \autoref{subsec:DTs}.

Unfortunately, sometimes there are real-world considerations that conflict with the goals of data mining; sometimes the privacy of the people being data mined needs to be considered. Privacy plays a role in scenarios ranging from government projects like a census, to businesses collecting information about their present and future customers, to health organizations analyzing illnesses or hospitals admissions \cite{Mohammed2015}. In some of these cases, the government may mandate certain privacy protections; in other cases, businesses may want to encourage customers to provide more personal data by promising to protect their privacy. These privacy considerations necessitate that the original data be modified in some way, or that the output of data mining algorithms be modified, while hopefully not ruining the information that can be discovered from the data \cite{Brankovic1999}.

Many strategies for protecting privacy have been explored in recent years \cite{Adam1989,Aggarwal2008c,Fung2010}. Noise addition \cite{Agrawal2000,Islam2011}, $k$-anonymity  \cite{Sweeney2002} and $l$-diversity \cite{Machanavajjhala2007} are several such strategies, however these strategies provide privacy in a way that is not ``guarantee-able''; that is, they do not provide privacy in a mathematically rigorous way that guarantees that every person with sensitive information in a dataset will have their privacy protected. A recent example of this phenomena was seen in Australia, were the government released a health dataset describing one billion insurance claims since 1984 to researchers. The privacy of the people in the dataset was protected using ``a suite of confidentiality measures including encryption, perturbation and exclusion of rare events'' \cite{Cowan2016}. This approach received wide criticism for not being rigorous enough, including from Australia's Privacy Commissioner, Timothy Pilgrim \cite{Cowan2016a}. 

Differential privacy, proposed in 2006, is a privacy-preserving strategy that rectifies the weaknesses of other strategies by guaranteeing that nothing can be learned about any individual in the dataset, that could not have been learned even if the individual was not in the dataset at all \cite{Dwork2006,Dwork2014}. It is this definition of privacy preservation, and how it can be applied to decision trees, that we focus on in this survey. We introduce differential privacy in more detail in \autoref{subsec:DP}. Differential privacy has become the de-facto privacy standard around the world in recent years, with the U.S. Census Bureau using it in their \emph{Longitudinal Employer-Household Dynamics Program} in 2008 \cite{Machanavajjhala2008}, and the technology company \emph{Apple} implementing differential privacy in their latest operating systems and applications \cite{Greenberg2016}.

Several previous works have surveyed the conflict between data mining and differential privacy \cite{Ji2014,Sarwate2013}. \citeN{Sarwate2013} provided a general overview of machine learning with differential privacy, briefly discussing classification, regression, dimensionality reduction, time series, filtering, and a suite of some of the basic ``building blocks'' of differential privacy, such as the differences between input, output and objective perturbation, the Exponential mechanism, and robust differentially private statistics. \citeN{Ji2014} focused more on specific differentially private data mining algorithms, providing an overview of the work done with naive Bayes models, linear regression, linear SVM's, logistic regression, kernel SVM's, decision trees, online convex programming, $k$-means, feature selection, PCA, and statistical estimators. Aside from the above two surveys, \citeN{Yang2012} provide some additional resources and references in their short workshop paper. To the best of our knowledge, these are the only surveys that previously explored the intersection between data mining and differential privacy. For a broader understanding of the theory underpinning differential privacy, we recommend \citeN{Li2016}'s and \citeN{Zhu2017}'s books on the subject.

Unlike previous surveys, we choose to focus on the intersection between differential privacy and decision trees, and how the structures of decision tree algorithms interact with differentially private querying. We examine the intricacies of differentially private decision tree algorithms in much greater detail than any previous work, and interrogate the design decisions that need to be made in order to make an algorithm differentially private without destroying its utility. By performing a deep dive on how a particular type of machine learning algorithm can be made differentially private, we hope to provide the reader with the necessary skills to design their own high-utility differentially private algorithms.

We begin by providing the necessary background knowledge for both differential privacy and decision tree classification in \autoref{sec:prelims}. Then in \autoref{sec:main}, we break down each component of decision tree algorithms and discuss when and how the data can be queried using differential privacy. For each component, whether it is leaf nodes, non-leaf nodes, termination criteria, pruning, or multiple trees, the decision to query the data comes with a privacy cost. These privacy costs add up, and ultimately dictate the utility of the decision tree. Great care must therefore be taken when deciding which components are data-based, and which components can be navigated without querying the data. \autoref{sec:together} looks at proposed implementations of differentially private decision tree algorithms, and compares the advantages and disadvantages of each algorithm. \autoref{sec:future} looks forward into the future, and recommends directions of future research that may push differentially private decision trees even further. \autoref{sec:publish} provides a different perspective of privacy preservation, looking at how non-private decision trees can be used to facilitate the publication of differentially private raw data. \autoref{sec:budget} discusses how the fundamental parameter that differentially private decision trees rely on, $\beta$, can be budgeted in practical real-world scenarios. We then conclude the survey in \autoref{sec:conclusion}.

\section{Preliminaries} \label{sec:prelims}

\subsection{Differential Privacy} \label{subsec:DP}

Differential privacy is a definition that makes a promise to each individual who has personal data in a dataset: ``You will not be affected, adversely or otherwise, by allowing your data to be used in any analysis of the data, no matter what other analyses, datasets, or information sources are available'' \cite{Dwork2014}. It does not promise that nothing will be discovered about the individual, just that whatever is discovered about them would have been discovered even if their data was not in the dataset at all. It also promises that any supplementary data a malicious user might have about the individual is irrelevant; the attacker can know any amount of information about an individual, and even know every single other data point in the dataset, and still not be able to detect the presence of the targeted individual. This is not a 100\% guarantee, but instead a very high probability guarantee. The exact probability is determined by a parameter $\epsilon$. The value of $\epsilon$ is chosen by the data curator in the ``curator model'' (where a trusted curator wishes to safely output information about the data) and by either the curator or each individual in the ``local model'' (where each person's data is anonymized before being received by the curator). The topic surveyed in this paper, differentially private decision trees, uses the curator model. The smaller $\epsilon$ is, the higher the privacy guarantee. 

\begin{figure}[t]
	\centering
	\includegraphics[width=4.5in]{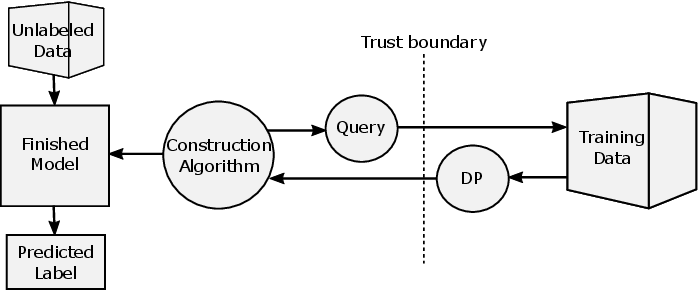}
	\caption{High-level representation of the user's interface with private data, using differential privacy (DP).}
	\label{fig:pipeline}
\end{figure}

\autoref{fig:pipeline} presents a high-level view of the pipeline used by differentially private data mining algorithms; an algorithm submits a query to the dataset, the dataset calculates the answer to the query, and then a differentially private mechanism modifies the answer in a way that protects the privacy of every individual person in the dataset.

We briefly define differential privacy, and some differentially private mechanisms that decision trees can take advantage of. An in-depth description of the many facets of differential privacy is provided by \citeN{Dwork2014}. We write the following definitions in terms of some query (i.e., function) $f$ submitted to a dataset $x$ describing $n$ records (i.e., individuals) from a universe $D$. We compare $x$ to a neighboring dataset $y$, where ``neighbor'' describes how many records differ between $x$ and $y$ (i.e., their Hamming distance in terms of records).

\begin{theorem}[Differential Privacy \cite{Dwork2006}] \label{def:DP}
A non-deterministic function $M$ (i.e. a function with a randomized component) is $\epsilon$-differentially private if for all outputs $g\subseteq Range(M)$ and for all data $x,y\in D^n$ such that $||x-y||_1\leq 1$: 
\begin{equation}
Pr(M(x)=g) \leq e^\epsilon \times Pr(M(y)=g) \enspace .
\end{equation}
Function $M$ will often be ``wrapper'' function around a deterministic function $f$.
\end{theorem}

The parameter $\epsilon$ can be considered as a ``cost'', with multiple costs summing together, described below:
\begin{theorem}[Composition \cite{McSherry2007}] \label{def:composition}
The application of all functions $\{M_i(x)\}$, each satisfying $\epsilon_i$-differential privacy, satisfies $\sum\limits_i {\epsilon_i}$-differential privacy.
\end{theorem}
If the same query (i.e. function) is submitted to multiple subsets of the data, with no overlapping records, the costs do not need to be summed:
\begin{theorem}[Parallel Composition \cite{McSherry2009}] \label{def:parallel}
For disjoint subsets $x_i\subset x$, let function $M(x_i)$ satisfy $\epsilon$-differential privacy; then applying all functions $\{M(x_i)\}$ still satisfies $\epsilon$-differential privacy.
\end{theorem}

Differential privacy is a definition, not an algorithm; mechanisms need to be designed that allow a user to query data in a way that adheres to the definition. These mechanisms are often formulated as function $M$, which takes a non-private function $f$ as input and converts it into a differentially private function. A popular mechanism for outputting a real number is to use the following Laplace mechanism, and a discrete value can be outputted with the Exponential mechanism:
\begin{theorem}[Laplace Mechanism \cite{Dwork2006a}] \label{def:laplace}
A query $M$ satisfies $\epsilon$-differential privacy if it outputs $M(x)=f(x)+L$, where $f:f(x)\rightarrow\mathbb{R}$ and $L\sim Lap(\Delta(f)/\epsilon)$ is an i.i.d. random variable drawn from the Laplace distribution with mean 0 and scale $\Delta(f)/\epsilon$. We shorten $L\sim Lap(\Delta(f)/\epsilon)$ to $Lap(\Delta(f)/\epsilon)$ when our meaning is clear from context. $\Delta$ is described below.
\end{theorem}
\begin{theorem}[Exponential Mechanism \cite{McSherry2007}]\label{def:expo}
Using a scoring function $u:u(z,x)\rightarrow\mathbb{R}$ where $u$ has a higher value for more preferable outputs $z\in Z$, a query $M$ satisfies $\epsilon$-differential privacy if it outputs $z$ with probability proportional to $\exp{(\frac{\epsilon u(z,x)}{2\Delta(u)})}$. That is,
\begin{equation}\label{eq:expo}
Pr(M(x)=z) \propto \exp{\left(\frac{\epsilon\times u(z,x)}{2\Delta(u)}\right)} \enspace .
\end{equation}
\end{theorem}

The above mechanisms add noise to the output that is scaled to the ``sensitivity'' $\Delta$ of the query, which is defined as the maximum amount that the output could change by if $x$ had one record added or removed. Other definitions of sensitivity exist \cite{Nissim2007a}, but for now we limit the discussion to global sensitivity:
\begin{theorem}[Global Sensitivity \cite{Dwork2006a}] \label{def:sensitivity}
A query $f$ has global sensitivity $\Delta(f)$, where:
\begin{equation}\label{eq:sensitivity}
\Delta(f) = \max_{x,y:||x-y||_1\leq 1}||f(x)-f(y)||_1 \enspace .
\end{equation}
\end{theorem}

\subsubsection{Factors that Affect Differentially Private Algorithms} \label{subsec:DP_factors}

The main factors that need to be considered when designing a differentially private machine learning algorithm are:
\begin{itemize}
	\item How large of a privacy budget $\beta$ the data curator is providing the user with. The total budget dictates the overall constraints put on the data mining algorithm. We discuss the range of sizes $\beta$ can have in real-world scenarios in \autoref{sec:budget}.
	\item The number of times the data needs to be queried. The more queries that the algorithm needs, the more the total privacy budget $\beta$ needs to be divided up to pay for them all. This reduces how large $\epsilon$ can be for each query, and the smaller $\epsilon$ is, the noisier the outputs of the queries will be. Limiting the number of required queries is discussed in \autoref{subsec:nonleaf}, \autoref{subsec:termination} and \autoref{subsec:efficiency}.
	\item The sensitivity $\Delta$ of the queries, which is influenced by the range and distribution (i.e. shape) of the data \cite{Hay2016}. Sometimes a query that performs well in a \emph{non-private} setting becomes unusable due to how sensitive it is to individual records, leading to overwhelming noise being added to the output. Instead, traditionally sub-optimal queries in non-private settings can be preferable in the private setting if they have low sensitivity. We explore this phenomena in \autoref{subsec:nonleaf}.
		\item The size (or scale) $n$ of dataset $x$ also plays an important role. The amount of noise that must be added to enforce differential privacy is independent of $n$, so the larger $n$ is, the smaller the relative amount of noise becomes.
\end{itemize}

\subsection{Conventional, Non-private Decision Trees}\label{subsec:DTs}

Decision trees are a non-parametric supervised learning method used for classification and regression \cite{Han2006a}. They make no assumptions about the distribution of the underlying data, and are trained on labeled data to correctly classify previously unseen data. They have several advantages over other kinds of supervised learning methods that make them appealing to data scientists. Some advantages include their:
\begin{itemize}
	\item high human interpretability \cite{Huysmans2011,Letham2013}
	\item non-parametric design \cite{Murphy2012}
	\item relatively low computational cost \cite{Han2006a}
	\item ability to discover non-linear relationships among the attributes \cite{Han2006a}
	\item resilience to missing values \cite{Quinlan1993}
	\item ability to handle both continuous and discrete data \cite{Quinlan1996}
	\item ability to handle non-binary labels \cite{Quinlan1996}
\end{itemize}
Their main disadvantages -- tendency to over-fit the data and instability to small changes in the data -- are minimized by limiting how deep the trees can grow, pruning away untrustworthy leaf nodes, building an ensemble of trees instead of just one, and using bootstrapped data samples in each tree \cite{Breiman2001}. Other disadvantages can make them unsuitable in some scenarios, such their difficulty in finding non-orthogonal decision boundaries or express XOR relationships \cite{Murphy2012}. Despite some disadvantages, their advantages make them an apt choice in a wide variety of data mining scenarios. This paper focuses on classification trees, with discrete outcome variables.

The ``untrustworthiness'' of a node can be defined in a variety of ways \cite{Letham2013}, but the most common methods involve ``support'' and ``confidence''. Support refers to the size of the data subset (i.e., the number of records) in the node. Confidence, also known as ``purity'' \cite{Kotsogiannis2017}, refers to the homogeneity of the class labels in the node; that is, the percentage of records with the most common class label in the node. For most intents and purposes, the class is the same as any other attribute of the dataset, except that it has been specially chosen to act as a way of classifying records. The class $C$ has multiple possible values $c\in C$, often called ``labels'', with each record having one or more labels. The larger the support and confidence, the more a node is normally trusted. Tangentially, a node's ``interestingness'' is another active field of research, which aims to describe how much useful information is contained in a node \cite{Geng2006,Guillet2007}.

\begin{figure}[t]
	\centering
	\includegraphics[width=3.0in]{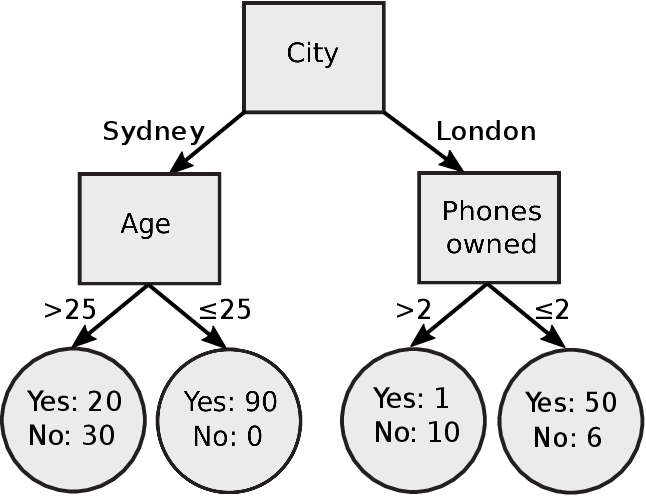}
	\caption{An example of a decision tree, with a depth of $d=3$.}
	\label{fig:tree}
\end{figure}

Fig.~\ref{fig:tree} is an example of a decision tree. Put briefly, a decision tree is an acyclic directed graph, built using top-down recursive partitioning of the dataset \cite{Han2006a}. The records in the dataset are recursively divided into subsets using ``tests'' in each node of the tree. The test checks the value each record has for a chosen attribute, sending records to the child node that matches the value they have. An attribute is chosen in each node based on that attribute's ability to split records into subsets that have as homogeneous a class label as possible. For continuous attributes, the best splitting value is found, splitting the records into two child nodes based on whether records' value is higher or lower than the splitting value. For discrete attributes, either a child node is created for every value, or a single value is chosen to split the records based on having that value or not having that value.

Once the tree has recursively divided the data the maximum number of times (i.e., reached the maximum tree depth) or the nodes do not have enough records to reliably classify records outside the training data (i.e., over-fitting becomes likely), tree-building is stopped. The resulting tree can then undergo a small amount of back-propagation in the form of ``pruning'', where untrustworthy leaf nodes are removed from the bottom of the tree \cite{Han2006a}. The tree is then ready for classifying, whereby unseen records can be filtered through the tree based on the tests in each node, finally ending up in the only leaf node for which it obeys all the preceding tests. The majority (i.e., most common) class label out of the training records in that leaf node is then the predicted class label of the new record. The path a record took from the root node to a leaf node is sometimes called the root-to-leaf path, or the ``rule'' that the record obeys \cite{Menardi2014}. Each root-to-leaf path in a decision tree is a different rule, and the set of rules can be considered a summary of the information conveyed by the dataset.

If the tree is built greedily (explained in \autoref{subsec:greedy}), local optima are a possibility, which could lead to parts of the tree being of much lower quality than other parts of the tree. Small fluctuations in the training data can also cause different attributes to be chosen in early nodes, cascading into a completely different tree structure from what otherwise might be built. To avoid these problems, a forest (i.e., an ensemble or collection) of trees is built, with unseen records being filtered through every tree, and the most common prediction out of all the trees being used as the final predicted class label (a process commonly referred to as ``voting'' \cite{Han2006a}).

Within this basic tree-building process, there are two fundamentally different approaches that can be taken: greedy approaches and random approaches. We briefly discuss the differences below. For a more thorough explanation of decision trees, we refer the reader to \citeN{Han2006a}.

\subsubsection{Greedy Decision Trees} \label{subsec:greedy}

The process of top-down decision-making is sometimes referred to as ``greedy'' decision-making, due to its strategy of making the optimal decision for the immediate short-term situation, with no regard for long-term consequences. While clearly not perfect, considering all possible eventualities of a system is usually NP-hard, and greedy heuristics perform well in practice \cite{Han2006a}. Greedy decision trees use this heuristic strategy, where an objective function is maximized in each node of the tree to decide how to split the node into child nodes. Many different splitting functions have been proposed for decision trees in the past, such as information gain \cite{Quinlan1993}, gain ratio \cite{Quinlan1996}, and gini index \cite{Breiman1984}, and they all aim to maximize the discriminatory power of the child nodes in as unbiased \cite{Quinlan1996} a way as possible. We briefly describe one such splitting function, information gain, as an example below:

\paragraph{Information Gain}

Informally, information gain uses Shannon's definition of information as being the opposite of entropy \cite{Shannon1949}; reducing entropy means gaining information. More formally, information gain can be expressed as the entropy of the current node $i$ minus the total entropy of the child nodes $J$ created by splitting $i$ with each value of an attribute $A$. Since the aim is to find the attribute that best reduces entropy, and the entropy of $i$ is the same regardless of which attribute is selected to make child nodes $J$, only the entropy of $J$ needs to be calculated in order to find the attribute with the largest information gain. The attribute that will best split $i$ is the attribute that minimizes:
\begin{equation}\label{eq:infogain}
InfoGain(x_i,A) = -\sum_{v\in A} \left(\frac{n_i^v}{n_i}\sum_{c\in C} \frac{n_i^{v,c}}{n_i^v}\log_2\frac{n_i^{v,c}}{n_i^v}\right) \enspace ,
\end{equation}
where $x_i$ is the subset of $x$ in node $i$, $n_i^v$ is the number of records in $i$ with value $v\in A$, and similarly for $n_i^{v,c}$ with class label $c\in C$.

Examples of greedy decision tree algorithms are ID3 \cite{Quinlan1986} (which was developed further into C4.5 \cite{Quinlan1993}), CART \cite{Breiman1984} and \citeN{Hothorn2006}'s unbiased regression tree algorithm. These algorithms built a single tree, and were extended in later years to build an ensemble of trees. Random Forest \cite{Breiman2001} is one such extension, which selects a random subset of records and a random subset of attributes to build each greedy decision tree with, often 100 or more times. The random selection of records and attributes adds diversity to the decision trees, helping prevent the greedy splitting function from getting trapped in local optima. It also prevents the final decision forest from being over-reliant on specific records, which can otherwise lead to fragile tree structures.

\subsubsection{Random Decision Trees} \label{subsec:random}

\citeN{Breiman2001} took advantage of randomness to improve the performance of greedy decision trees when he invented Random Forest. This idea has since been taken much further, with greedy heuristics being removed altogether in some cases \cite{Fan2003,Geurts2006}. For these random decision trees, nodes are instead split by randomly selecting an attribute. In the case of continuous attributes, the splitting point can also be uniformly randomly selected from the attribute's range. This approach works extremely poorly for a single tree; it only performs well when many random trees are used in combination. The computational cost of this approach is also much lower than an ensemble of greedy decision trees, since it avoids calculating the output of an objective function for every attribute in every node of every tree. 

When multiple trees are used in combination like this, it is known as an ensemble classifier \cite{Maudes2012}. When explicitly referring to a complete ensemble classifier, we use the phrase ``decision forest'' to distinguish from situations where either single trees, or trees inside a forest, are being discussed.

\subsubsection{Factors that Affect Tree Classification}

The main factors to consider when designing a decision tree algorithm are:
\begin{itemize}
	\item What kinds of dataset properties the algorithm is catered towards, such as if it can handle discrete attributes, continuous attributes, or both. The dimensionality of the dataset also plays a large role, mostly in terms of how the number of attributes $m$ affects tree depth, discussed in \autoref{subsec:termination}. The number of records $n$ also plays a role in defining termination criteria, but predominately due to the requirements of differential privacy, as mentioned in \autoref{subsec:DP_factors}. The overall role of the data is discussed in \autoref{subsec:data}.
	\item What splitting function to use (including random selection). We explore the effect of this decision in \autoref{subsec:nonleaf}.
	\item What termination criteria to use. We discuss four different types of terminations, and several examples of each, in \autoref{subsec:termination}.
	\item Whether to include a pruning step, and what pruning would be most appropriate if so. We discuss pruning in \autoref{subsec:pruning}.
	\item Whether to build multiple trees and use each as part of a larger ensemble, and how many trees to build if so. We explore these ideas in \autoref{subsec:forest}.
\end{itemize}

\section{Differentially Private Decision Tree Classification} \label{sec:main}

When outputting a decision tree, privacy is leaked via the information describing each node. At its most basic level a decision tree algorithm is deciding which attribute to split each node with (e.g. with a splitting function; see \autoref{eq:infogain}), and this decision is dictated by the data in the node. Once the tree has finished being built, the leaf nodes can output some information about the class counts, which is also dictated by the data in the nodes (see \autoref{subsec:DTs}). Since these decisions and outputs are directly based on the data, differential privacy states that releasing the information can be a breach of privacy. These potential breaches are what a differentially private decision tree algorithm aims to prevent.

All the algorithms we discuss in this survey achieve the same basic goal of outputting information about nodes in a differentially private way, but do so with a wide variety of strategies. Some algorithms intelligently find patterns in the data, providing the user with knowledge and not just a black-box classifier. Other algorithms sacrifice everything that is not absolutely necessary in order to build an accurate classifier, submitting as few queries as possible. These two extremes, and everything in between, offer different trade-offs between knowledge, accuracy, noise, and privacy costs. Cleverly using the privacy budget is imperative when designing an effective differentially private decision tree classifier.

\subsection{When is the data needed?} \label{subsec:data}

Part of the privacy budget $\beta$ needs to be spent whenever the private data is queried. This means it is important to identify when exactly the data \emph{needs} to be queried in order to build a decision tree. The less times the data needs to be queried, the less budget we need to spend, or the more budget we can spend per query. Sometimes the data will not necessarily \emph{need} to be queried, but doing so will still be worth the privacy cost due to how much better the classifier performs because of it. Spending part of the budget on optional queries is explored in \autoref{subsec:efficiency}. 

Which queries should be considered ``compulsory'' depends on one major characteristic of the tree-building algorithm: whether it builds the tree greedily or randomly. By ``greedily'', we refer to an algorithm using an objective function locally in each node; heuristically finding the best attribute to divide the local subset of data contained in a node (see \autoref{subsec:greedy}). No matter what objective function is used (be it information gain, gini index, or any other), it requires querying the local data at least once. We refer to these as ``non-leaf queries'', since they are performed in each non-leaf node in a tree, and explore them in detail in \autoref{subsec:nonleaf}. The other compulsory query for greedy trees is the same as the only compulsory query for random trees. By ``random'' we refer to algorithms that do not use an objective function in each node, and instead pick attributes randomly (thus having no need for the data; see \autoref{subsec:random}). The compulsory query that both categories of decision trees share is a query that enables the prediction of class labels for unseen, unlabeled data. In almost all cases, this takes the form of querying the distribution of class counts in each leaf node, with the most common (i.e., majority) class label being used as the predicted label for future records (see \autoref{subsec:DTs}). We refer to these queries as ``leaf queries'' and discuss them in \autoref{subsec:leaf}.

It is worth mentioning that in almost all scenarios involving differential privacy, the attribute schema (that is, the domains of the attributes) is considered to be public knowledge \cite{Blum2005,Dwork2014,Friedman2010,Jagannathan2012,Rana2016}. This information can be dependent on the data in the sense that the domain of a discrete attribute is the set of discrete values in the dataset $x$, but it can also contain values that are possible in the universe $D$ but lack any actual examples in $x$. This is the recommended approach for continuous attributes \cite{Dwork2014}; defining the lower and upper bounds of a continuous attribute using the minimum and maximum values found in $x$ can be a breach of privacy, and so instead it is better to define the bounds using the universe $D$. For example in \autoref{fig:tree}, the domain of the ``Age'' attribute might be $[0,120]$, even if the oldest person in $x$ is $96$. This approach can be used for discrete attributes as well, though defining the domain of an attribute like ``City'' as all possible cities on Earth is likely to make the attribute untenable when trying to query it. It is therefore the responsibility of the data curator to provide a reasonable attribute schema to the users wishing to query the data. It is possible for users to use their best estimations to create a reasonable schema of their own, but this is susceptible to practical issues such as value formatting and spelling. Without a public attribute schema, it is difficult to imagine how queries can even be submitted to a database, and thus this information is considered to be known by the user within the scope covered by this survey.

\subsection{Non-leaf Queries} \label{subsec:nonleaf}

Deciding which attribute to split a node with to optimize its ability to discriminate between class labels is at the core of a greedy decision tree algorithm. It is what \citeN{Blum2005} focused on during differential privacy's inception in 2005, where they proposed a simple proof-of-concept decision tree algorithm. This algorithm demonstrated how a traditional non-private algorithm could be converted to achieve differential privacy, by rephrasing the splitting function in terms of queries that could be made differentially private. Specifically, \citeN{Blum2005} made Information Gain (see \autoref{eq:infogain}) differentially private by breaking it down into two counting queries for each attribute $A$, and then making the counting queries differentially private with the Laplace mechanism (\autoref{subsec:DP}):
\begin{description}
	\item[Blum et al.'s Query 1] $n_i^{v,c}+Lap(1/\epsilon); \forall c\in C, \forall v\in A$ and
	\item[Blum et al.'s Query 2] $n_i^v+Lap(1/\epsilon); \forall v\in A$.
\end{description}
where $n_i$ is the number of records (i.e., support) in node $i$, $n_i^v$ (and $n_i^{v,c}$) is the number of records in $i$ with value $v$ (and class label $c$), and $\epsilon$ is the privacy budget spent on that query. With these two queries, \autoref{eq:infogain} can then be calculated client-side. Differentially private outputs are immune to post-processing privacy breaches, so nothing more needs to be done to satisfy differential privacy in the non-leaf nodes.

It is, however, a costly approach to take when considering the privacy budget. If we consider the user's total privacy budget to be $\beta$, then the two queries listed above can only receive a small fraction of $\beta$ each. When querying any given attribute $A$ with the above queries, the counts for each combination of value $v$ and label $c$ involves a disjoint subset of $x_i$ (i.e., the data in node $i$), allowing \emph{Query 1} to be composed in parallel $\forall c\in C, \forall v\in A$ (\autoref{def:parallel}). \emph{Query 2} uses the same data as the first query though, and so its privacy cost is summed with the cost of the first query (\autoref{def:composition}). This then needs to be repeated for every attribute $A$, at every level of the tree (sibling nodes contain disjoint subsets of $x$ and can be composed in parallel). When all is said and done, each query in the algorithm has a fraction of the total privacy budget $\beta$ equal to:
\[
\epsilon = \frac{\beta}{2md} \enspace ,
\]
where $m$ is the number of attributes and $d$ is the tree depth (including the root and leaf levels)\footnote{Note that strictly speaking, the leaf nodes only require one query; the total class counts in each leaf node. We discuss leaf queries in \autoref{subsec:leaf}}. This expenditure of the budget is inefficient for several reasons. One inefficiency can be immediately seen when looking at the two queries listed above; the second query could be calculated by simply summing the outputs from the first query! That is, $n_i^v = \sum_{c\in C} n_i^{v,c}$, allowing the user to skip the second query entirely. This would involve summing the $Lap(1/\epsilon)$ noise added to each $n_i^{v,c}$ count, but this actually results in lower noise due to a lot of the noise ``canceling out'' \cite{Fletcher2015c}. We explore the idea of summed noise canceling itself out in \autoref{subsec:pruning}, and further budget-saving strategies in \autoref{subsec:efficiency}.

Converting a splitting function into differentially private counting queries is fortunately not the only approach one can take when building a greedy tree. Rather than submitting counting queries via the Laplace mechanism (\autoref{def:laplace}), one of the major advances \citeN{Friedman2010} made over the preliminary work done by \citeN{Blum2005} was to submit a single query to find the best splitting attribute in each node via the Exponential mechanism (\autoref{def:expo}).

\citeN{Friedman2010} proposed building a decision tree by querying the dataset twice at each node:
\begin{description}
	\item[Friedman and Schuster's Query 1] first, get a count of how many records are in the node, followed by
	\item[Friedman and Schuster's Query 2] find the best attribute to split the records with.
\end{description}
These two non-leaf queries are also used by \citeN{Patil2014}, \citeN{Fletcher2015b} and \citeN{Rana2016} in their greedy tree algorithms, which we discuss later in this section. For any particular level (i.e., depth) of the decision tree, all the nodes contain disjoint subsets of the dataset, allowing for parallel composition to be used (\autoref{def:parallel}). The privacy budget spent on the two parallel queries at each level of the tree can then be summed using composition (\autoref{def:composition}). This means that for a given total privacy budget $\beta$, each query submitted to the dataset has a portion of the budget equal to 
\[
\epsilon=\frac{\beta}{2d} \enspace ,
\]
an $m$-fold improvement over the proof-of-concept algorithm offered by \citeN{Blum2005}. Note that the first query (i.e., the support of a node) is only used as part of the algorithm's termination criteria (\autoref{subsec:termination}) and for pruning (\autoref{subsec:pruning}), and is not discussed here.

\begin{figure}[t]
	\centering
	{\footnotesize \def\svgwidth{4.4in} 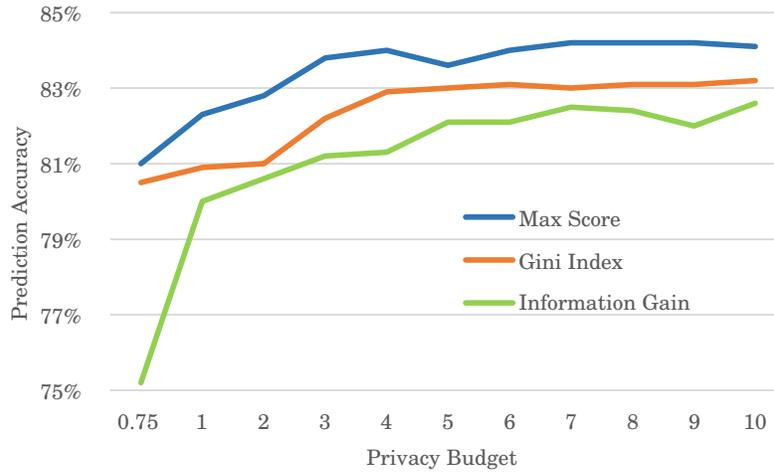}
	\caption{The prediction accuracy achieved when building Friedman and Schuster's decision tree with three different splitting criteria, using the Adult dataset. Based on results presented in Friedman and Schuster [2010].}
	\label{fig:splitting}
\end{figure}

Using the Exponential mechanism, the best attribute to split records in a node with can be outputted with high probability. The probability of any given output being chosen by the Exponential mechanism depends on the scoring function, which in this case is the splitting function. Common splitting functions are information gain \cite{Quinlan1993}, gain ratio \cite{Quinlan1996}, and gini index \cite{Breiman1984}. Friedman and Schuster analyzed each of these; not in terms of utility performance (which has been done many times in non-private scenarios \cite{Li2002,Islam2012,Katz2012}), but in terms of how sensitive they were to individual records. The sensitivity of a function directly impacts how much noise needs to be added to achieve differential privacy. In their experiments Friedman and Schuster found that due to its low sensitivity, using max operator \cite{Breiman1984} (see \autoref{subsec:greedy}) as their splitting function achieved the highest prediction accuracy, despite it being the least sophisticated of the tested functions. Their findings can be see in \autoref{fig:splitting}. This demonstrates that taking a function that performs well in \emph{non-private} scenarios and making it differentially private does not necessarily continue to perform well in \emph{private} scenarios. Gini Index also achieved competitive results, having much lower sensitivity than Information Gain.

Interestingly, \citeN{Friedman2010} were unable to test Gain Ratio due to its erratic behavior when the denominator of the ratio approaches zero. Quinlan's C4.5 algorithm \cite{Quinlan1996} solved this problem with additional heuristics, but ultimately the sensitivity of Gain Ratio cannot be bounded and thus cannot be implemented with the Laplace mechanism or the Exponential mechanism. 

\subsubsection{Reducing the Sensitivity of Splitting Functions} \label{subsec:split_sens}

In an effort to reduce the amount of noise induced in the tree-building process, \citeN{Fletcher2015b} explored reducing the sensitivity of the splitting functions. Instead of using the theoretical worst-case scenario to calculate the \emph{global} sensitivity of a function, they took advantage of the (otherwise underused; see \autoref{subsec:efficiency}) first query by Friedman and Schuster and calculated the \emph{local} sensitivity of the gini index in the node being split. Friedman and Schuster's first query returns the support of the node, and this (now public) information can be used by the user to propose a better-bounded sensitivity to the server. The authors proved that the sensitivity of the gini index could be reduced from $\Delta=0.5$ to 
\begin{equation} \label{eq:gini_sens}
\Delta=1-\left(\frac{n_i}{n_i+1}\right)^2-\left(\frac{1}{n_i+1}\right)^2 \enspace ,
\end{equation}
where $n_i$ is the support of the node being split, node $i$. Reductions in sensitivity such as this one directly reduce the noise induced by mechanisms like the Laplace mechanism and the Exponential mechanism.

To be fully differentially private however, the sensitivity reduction proposed by \citeN{Fletcher2015b} needs to be ``smoothed'' out to account for all neighboring datasets $y$, including datasets at a distance $k>1$ from $x$. In its proposed form, the sensitivity that the user calculates using \autoref{eq:gini_sens} is actually noisy, since $n_i$ is noisy. The server might therefore reject the proposed sensitivity if it is too small, since it would not properly provide differential privacy. A rough ``buffer'' could be added to the estimated sensitivity to make it more likely to be accepted by the server (a similar strategy is used by \citeN{Rana2016} when calculating continuous splitting points, discussed in \autoref{subsec:continuous}), but ultimately this is not mathematically rigorous enough for differential privacy. Conversely, if the exact sensitivity is outputted to the user and only neighbors $y$ where $k=1$ are considered, the sensitivity itself can leak private information \cite{Nissim2007a}. 

Smooth sensitivity, proposed by \citeN{Nissim2007a}, solves this problem with \citeN{Fletcher2015c}'s local sensitivity by taking it one step further. \citeN{Nissim2007a} propose an exact calculation of the local sensitivity that is strongly differentially private. Essentially, smooth sensitivity finds the maximum amount that a function $f$'s output can change when $k$ records are added or removed, given the specific input data $x$. This differs from the traditional global sensitivity not only because it considers the actual data instead of the theoretical worst-case scenario, but also because it considers neighboring datasets $y$ that are more than one record away from $x$. Smooth sensitivity considers the most that the output of $f$ could change at any distance $k$, ``smoothly'' lowering the probabilistic weights of larger distances $k$. Put more formally, the (non-private) local sensitivity is:
\begin{definition}[Local Sensitivity \cite{Nissim2007a}]\label{def:local}
For $f:D^n \rightarrow \mathbb{R}^d$ where $n,d\in\mathbb{N}$, and data $x\in D^n$, the local sensitivity of $f$ at $x$ (with respect to the $\ell_1$ metric) is
\begin{equation}
LS_f(x) = \max_{y:||x-y||_1\leq 1}||f(x)-f(y)||_1 \enspace .
\end{equation}
\end{definition}
Local sensitivity can then be extended to any distance $k$ and smoothed:
\begin{definition}[Smooth Sensitivity \cite{Nissim2007a}]\label{def:smooth}
The local sensitivity of $f$, with distance $k$ between datasets $x$ and $y$, is 
\begin{equation}
S^k(x) = \max_{y:||x-y||_1\leq k} LS_f(y) \enspace .
\end{equation}
The smooth sensitivity of $f$ can now be expressed using $S^k(x)$:
\begin{equation}\label{eq:smooth}
S^*(f, x) = \max_{k=0,1,...,n} e^{-k\epsilon}S^k(x)
\end{equation}
where $\epsilon$ is the privacy budget of $f$.
\end{definition}

Smooth sensitivity is successfully used in a random decision tree algorithm proposed by \citeN{Fletcher2017}, discussed later in \autoref{subsec:leaf}.

Interestingly, the inability to bound the sensitivity of Gain Ratio discovered by \citeN{Friedman2010} could also have been circumvented by using smooth sensitivity. In fact, using smooth sensitivity would have allowed all four splitting functions (including Gini Index) to add less noise than was added by Friedman and Schuster, provided that the functions are applied to the data using the Exponential mechanism. This would be an interesting direction to explore in future research; using the smooth sensitivity of each of the splitting functions instead of the global sensitivity. We discuss these possibilities in \autoref{sec:future}.

\subsubsection{Continuous Attributes} \label{subsec:continuous}

\citeN{Friedman2010} also conducted a preliminary exploration into using splitting functions on continuous attributes. Using an attribute value from among the records in a node is a breach of privacy, so another solution is needed. Friedman and Schuster propose using the Exponential mechanism to select the best \emph{range} of attribute values for splitting, where all the values in the range output the same score from the splitting function. A data-independent value can then be uniformly randomly selected from the chosen range. Unfortunately, this approach requires a far higher number of queries than simply selecting the best discrete attribute to split a node with, requiring the privacy budget to be divided such that 
\begin{equation} \label{eq:continuous}
\epsilon=\frac{\beta}{(2+n)d}
\end{equation}
where $n$ is the number of continuous (i.e., numerical) attributes. For even small numbers of continuous attributes, this is a substantial increase in the amount of noise added to each query. To the best of our knowledge, there has not been a more budget-efficient method proposed for selecting the splitting point of continuous attributes in a (strongly) differentially private way, other than discretizing the attribute first. This is a weakness that would need to be addressed before greedy heuristics such as splitting functions can be considered a viable way to build a differentially private classifier when continuous attributes are involved.

\citeN{Rana2016} explored one possible way of efficiently finding a continuous splitting point, but it required weakening the definition of differential privacy. For a given continuous attribute $x$, rather than spending a large amount of $\beta$ to find the optimal splitting value like Friedman and Schuster, the authors instead find the average $x$ value for all the records in that node that have class label $c_1$, and similarly for $c_2$. The chosen splitting value is defined as halfway between those two average values. While this approach does not maximize the discriminatory power of the child nodes, it increases the budget per query from Friedman and Schuster's $\epsilon=\frac{\beta}{(2+n)d}$ to 
\begin{equation} \label{eq:rana_budget}
\epsilon=\frac{\beta}{3d} \enspace ,
\end{equation}
thus reducing noise.

To make this query (weakly) differentially private, \citeN{Rana2016} add Laplace noise $Lap(\Delta/\epsilon)$, where the sensitivity $\Delta$ is based on the smaller of the two class counts, $\Delta=3/\min(n_{c_1},n_{c_2})$. Instead of using smooth sensitivity or a strictly-correct global sensitivity, Rana et al. opt to assume that the continuous attributes have a normal distribution and estimate the sensitivity within 3 standard deviations, which is adequate in 99.7\% of cases (assuming the assumption holds). While not strictly differentially private, Rana et al. keep in the spirit of their weakened definition of differential privacy (analyzed in detail in \autoref{subsec:forest}) and instead heuristically add an amount of noise that could be considered ``adequately private'' in most scenarios. The assumption they choose to make is somewhat self-defeating however, as it negates one of the well-known advantages of decision trees; their non-parametric independence from the underlying distribution of the data.

\subsection{Leaf Queries}\label{subsec:leaf}

The information required from the data in a leaf node is quite different from the information required in a non-leaf node. Rather than learning the best attribute to partition the data with, the purpose of a leaf node is to predict the class label of unlabeled records that are filtered to the leaf. A different query from the ones discussed in \autoref{subsec:nonleaf} is therefore required; one related to the class labels of the training records in the leaf node.

A counting query is the most straight-forward solution, and is the solution used by almost every differentially private decision tree algorithm that we know of \cite{Blum2005,Friedman2010,Jagannathan2012,Fletcher2015b,Fletcher2015c,Mohammed2015,Patil2014,Rana2016}. Conceptually, one can consider the query as a single query submitted in parallel to all $j$ leaf nodes, returning a histogram of the class counts in those nodes:
\begin{equation}\label{eq:leaf}
\left\{ n_j^c + Lap(1/\epsilon); \forall c\in C \right\} ; \forall j \enspace .
\end{equation}
Alternatively, one can recognize (in much the same way that \cite{Dwork2006a} originally discovered) that since each count in a histogram is disjoint from the other counts, each class count in each leaf is simply being queried in parallel. These counts are made differentially private by adding $Lap(1/\epsilon)$ (see \autoref{def:laplace}).

The size of $\epsilon$ in \autoref{eq:leaf} depends on how much of the privacy budget is remaining after querying the non-leaf nodes. This is where random decision trees, described in \autoref{subsec:random}, gain a unique advantage; they do not query the non-leaf nodes at all. \citeN{Jagannathan2012} were the first to investigate the idea of an ensemble of $\tau$ differentially private random decision trees, spending the budget allocated to each tree on one query: \autoref{eq:leaf}. In other words, each class count in each leaf node in each tree has $Lap(1/\epsilon)$ added to it, where
\begin{equation} \label{eq:jag_budget}
\epsilon = \frac{\beta}{\tau} \enspace .
\end{equation}

Deciding how large $\tau$ should be is discussed later in \autoref{subsec:forest}; suffice to say that Jagannthan et al. recommend $\tau=10$. This is close to the portion of the budget allocated to the leaf nodes of the greedy trees proposed by \citeN{Friedman2010} when using their recommended depth $d=5$ (assuming few continuous attributes; see \autoref{eq:continuous}) and \citeN{Rana2016} (see \autoref{eq:rana_budget}).

\begin{figure}[t]
	\centering
	{\footnotesize \def\svgwidth{4.4in} 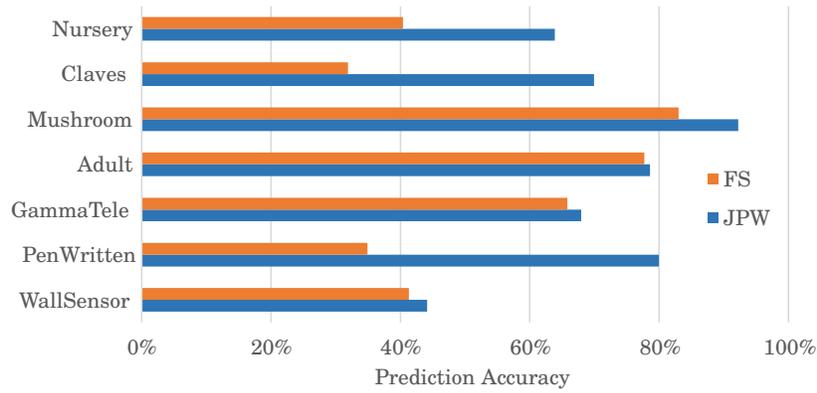}
	\caption{Comparing the prediction accuracy of Jagannathan et al.'s random decision forest algorithm (JPW) to Friedman and Schuster's greedy decision tree algorithm (FS) with seven datasets, when $\epsilon=1$. These two algorithms represent the most highly cited examples of their respective approaches: random vs. greedy. Based on results presented in Fletcher and Islam [2017].}
	\label{fig:jag_vs_fri}
\end{figure}

\citeN{Jagannathan2012}'s strategy proved successful; empirically, their random approach produces higher quality classifiers than \citeN{Friedman2010}'s greedy approach. \autoref{fig:jag_vs_fri} presents prediction accuracy comparisons with several datasets from the UCI Machine Learning Repository \cite{Dua2017}. Similarly to how the max operator splitting function outperformed more sophisticated functions in \autoref{subsec:nonleaf}, these results suggest that the benefits of using a splitting function that is more sophisticated than randomness may be lost when the output is so noisy. Instead, the ability to make multiple trees proves to be more valuable than building a single tree.

What random trees sacrifice is a lot of the discriminatory power created in each node by the splitting functions of greedy decision trees, and this is clearest when handling continuous attributes; rather than finding the optimal splitting point like \citeN{Friedman2010}, or an average value that tries to separate two binary class labels like \citeN{Rana2016}, Jagannathan et al. simply randomly choose a splitting point from the attribute's domain with uniform probability. This is the same strategy taken by traditional, non-private random decision trees \cite{Fan2003,Geurts2006}. Despite the average random node being less discriminatory than the average greedy node, the overall prediction accuracy of an ensemble of random trees has been shown to be very similar to the accuracy of an ensemble of greedy trees \cite{Geurts2006}.

\subsubsection{Using Smooth Sensitivity} \label{subsec:smooth}

To the best of our knowledge, the only differentially private decision tree algorithm to take advantage of smooth sensitivity \cite{Nissim2007a} is \citeN{Fletcher2017}. In order to use smooth sensitivity in the leaf nodes of a tree, the authors propose using a different query to the one presented in \autoref{eq:leaf}. Smooth sensitivity cannot help with counting queries; no matter what the specific data in $x$ is, the output of a counting query can always change by one when one record is added or removed. 

\begin{figure}[t]
	\centering
	{\footnotesize \def\svgwidth{4.4in} 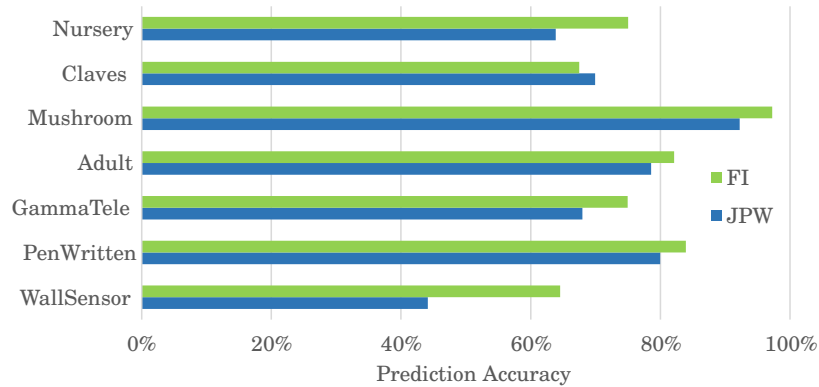}
	\caption{Comparing the prediction accuracy of Fletcher and Islam's random decision forest (FI) to Jagannathan et al.'s random decision forest (JPW) with seven datasets, when $\epsilon=1$. Based on results presented in  Fletcher and Islam [2017].}
	\label{fig:smooth_vs_jag}
\end{figure}

Similarly to how \citeN{Friedman2010} improved upon \citeN{Blum2005}'s original work by using the Exponential mechanism instead of the Laplace mechanism in the non-leaf nodes (see \autoref{subsec:nonleaf}), \citeN{Fletcher2017} do the same with the leaf nodes, outputting the majority class label with the Exponential mechanism. They observed that if the intent of the leaf nodes is \emph{only} to output a majority label, then the actual counts of the labels are largely irrelevant. Of course, this sacrifices the ability to learn any details about the leaf nodes for the sake of higher accuracy. Properties such as the support and confidence of the leaf nodes cannot be learned, as no numbers are outputted by the algorithm; only the majority labels. This makes it difficult to discover knowledge from the private data, such as the size or strength of the randomly generated rules. While not completely trustworthy due to the noise added, algorithms such as \cite{Jagannathan2012}'s random decision forest do allow for this kind of knowledge discovery. As is almost always the case when balancing the tension between privacy and utility however, the knowledge discovery comes at a cost; Jagannathan et al.'s algorithm is much less accurate at predicting the labels of unseen data. This can be empirically seen in \autoref{fig:smooth_vs_jag}. There is no free lunch when it comes to privacy \cite{Kifer2011}; it would be up to the user to decide if they want to sacrifice either knowledge discovery or prediction accuracy. These sorts of trade-offs are often present when designing a differentially private algorithm, making it important to have clearly defined objectives for what you want the algorithm to achieve.

To output the majority (i.e., most frequent) label in a leaf node with the Exponential mechanism, a scoring function (see \autoref{def:expo}) is required; \citeN{Fletcher2017} proposed a piecewise linear function to achieve this:
\begin{equation}\label{eq:score}
u(c, x_i) = 
	\begin{cases} 
      1 & c = \argmax_{c\in C} n_i^c \\
      0 & \mbox{otherwise}
   \end{cases}
	\enspace ,
\end{equation}
where $n_i^c$ is the number of occurrences of label $c\in C$ in node $i$. Each label will have a score of 1 if the label is the most frequently occurring label in the leaf, and 0 otherwise. The global sensitivity $\Delta(u)$ of this function is 1, since adding or removing one record could theoretically change a score from 1 to 0 or vice-versa. The authors demonstrate that the smooth sensitivity, however, is often much lower:
\begin{equation}\label{eq:score_smooth}
S^*(u, x_i) = e^{-j\epsilon}
\end{equation}
where $j$ is the difference between the most frequent and the second-most frequent labels in node $i$, $n_i^{c_1}-n_i^{c_2}$.

By taking advantage of smooth sensitivity, \citeN{Fletcher2017} were able to improve the accuracy of a random decision forest. Some empirical results are presented in \autoref{fig:smooth_vs_jag}, where the random decision forest proposed by \citeN{Fletcher2017} is compared to the one proposed by \citeN{Jagannathan2012}. 

The performance of smooth sensitivity here naturally prompts us to ask: what about greedy decision trees? What is the smooth sensitivity of splitting functions like information gain and gini index? Can the sensitivity of gain ratio now be calculated, something that \citeN{Friedman2010} demonstrated was not possible when using the global sensitivity? To the best of our knowledge, these questions remain open, and could be the key to allowing (strongly) differentially private greedy decision trees that match the performance of random decision trees, while also offering the opportunity for better knowledge discovery. We discuss these possibilities for the future in \autoref{sec:future}.

\subsection{Termination Criteria} \label{subsec:termination}

Knowing when to stop growing a decision tree is important in traditional non-private algorithms, but is even more important for a differentially private decision tree algorithm. This is due to two main reasons:

\begin{enumerate}
	\item When querying non-leaf nodes, only sibling nodes can be queried in parallel; different generations (i.e. levels) of nodes must have their privacy costs composed (see \autoref{subsec:DP}). The larger the depth, the more queries there are that need to be paid for.
	\item The data is split into smaller partitions as a tree gets deeper, and the smaller the class counts are in a leaf node, the larger the effect of the $Lap(1/\epsilon)$ noise added to them.
\end{enumerate}

The first reason affects only greedy trees, while the second affects both greedy trees and random trees. Throughout the literature, four main types of termination criteria are used when building either type of differentially private tree:
\begin{description}
	\item[Criteria 1] a user-defined criteria, such as maximum depth $d$ \cite{Friedman2010,Fletcher2015b,Patil2014} or minimum node support $n'$ \cite{Rana2016};
	\item[Criteria 2] a maximum depth based on the number of attributes $m$ \cite{Friedman2010,Jagannathan2012,Fletcher2017};
	\item[Criteria 3] defining a minimum node support $n'$ based on the ratio between the support and the added noise, and then either
	\begin{description}
		\item[a] using an estimation $\tilde{n_i}$ of node $i$'s support to decide whether to terminate \cite{Jagannathan2012,Fletcher2015c}; or
		\item[b] querying the actual support $n_i$ of node $i$ to more accurately determine whether to terminate \cite{Friedman2010}.
	\end{description}
\end{description}

Having the user define a maximum depth is the simplest approach, but it is also a naive approach that does not take into account any differences between datasets (unless the user employs some personal heuristics when making their decision). \citeN{Friedman2010} used this criteria in conjunction with two other termination criteria. Other researchers \cite{Fletcher2015b,Patil2014} use it as a placeholder criteria; their research does not focus on this component of the algorithm, but $d$ nonetheless needs to be defined in order to divide the budget $\beta$ among the queries. Many of the equations for dividing the budget $\beta$ in \autoref{subsec:nonleaf} refer to a maximum tree depth $d$, such as \autoref{eq:continuous} and \autoref{eq:rana_budget}.

\citeN{Rana2016} propose having the user define a minimum node support to act as the termination criteria, rather than a maximum tree depth. They define $n'$ as a some small fraction (such as $0.1\%$) of the total number of records $n$, where a node will not be split if any of the subsequent child nodes would have any class counts less than $n'$. This has a similar result to other minimum node support criteria (such as \citeN{Friedman2010}'s and \citeN{Fletcher2015c}'s, discussed later), except that Rana et al. rely on the user to manually define $n'$ rather than using a heuristic. Unfortunately, this approach leads to a problem: if we don't know how deep the tree will grow before we start querying the data, how do we decide what fraction of the privacy budget $\beta$ each query gets? For this reason, relying solely on $n'$ to terminate tree growth is not enough, and we include a maximum tree depth $d$ when discussing the privacy budgeting of Rana et al.'s approach (such as in \autoref{eq:rana_budget}). Note that using a minimum support criteria also requires querying the data in each node, which we discuss at the end of this section.

\citeN{Friedman2010}'s algorithm uses two more termination criteria, the next of which is based the number of attributes $m$. They only use it to avoid a logic error though; since discrete attributes can only be used once, the tree-building algorithm terminates if all discrete attributes have been tested (i.e. if $d=m$). For most datasets, $m$ will be larger than the user-defined maximum depth, leading to this termination criteria rarely being used.

The number of attributes can be used for more than just preventing logic errors though. As discussed in \autoref{subsec:random}, \citeN{Fan2003}'s combinatorial reasoning suggests that an ensemble of random trees has optimal diversity (and thus good discriminatory power) when each tree has a depth of $d=m/2$, and \citeN{Jagannathan2012} propose using the same idea with differentially private random trees. Because random trees do not query the data in non-leaf nodes, the portioning of the budget is independent of $d$ (see \autoref{eq:jag_budget}), allowing random trees to use whatever depth is considered optimal for \emph{non-private} random trees.

Note that using $d=m/2$ assumes that all the attributes can only be selected once in any root-to-leaf path. This is true for discrete attributes, but not continuous attributes. \citeN{Fletcher2017} updated \citeN{Fan2003}'s tree depth calculation to take into account continuous attributes, so that the number of different attributes tested in each root-to-leaf path remains optimal at $m/2$ different attributes tested. In other words, the authors proved that the optimal tree depth $d$ when using $m$ continuous attributes is
\[
d = \argmin_{\delta:X<m/2} \mathbf{E}[X|\delta]
\]
where $X$ is the number of attributes not tested on a root-to-leaf path of depth $\delta$,
\[
\mathbf{E}[X|\delta] = m\left(\frac{m-1}{m}\right)^\delta \enspace .
\]

There is one caveat with just using the same depth as non-private random trees though; if $m$ is large (and the tree is therefore deep), the tree might partition the data so many times that each leaf node ends up having a very small fraction of the total dataset size $n$. If the average support of the leaf nodes is small, adding $Lap(1/\epsilon)$ to each class count (see \autoref{subsec:nonleaf}) might completely overwhelm them with noise. \citeN{Jagannathan2012} therefore proposed a heuristic that uses the third type of termination criteria (i.e. estimating the nodes' support) and combined it with the $m/2$ criteria:
\begin{equation} \label{eq:jag_depth}
d = \min( m/2~,~\log_b n -1 )
\end{equation}
where $m$ is the number of attributes, $n$ is the number of records in the dataset, and $b$ is the average ``branching factor'' of the attributes. The branching factor of an attribute refers to how many child nodes will be created if a node is split using that attribute. This is the same as the domain size of the attribute if it uses discrete values, and is equal to $b=2$ if the attribute is continuous. Functionally, $\log_b n-1$ is setting the maximum depth to one less than the number of times $n$ can be partitioned evenly by each node (where each node has the average number of child nodes $b$) until the bottom nodes have support equal to one.

\autoref{eq:jag_depth} acknowledges that small class counts are more susceptible to noise than larger counts, but it does not take into account how much noise is being added. Depending on how small $\epsilon$ is, the class counts in the leaf nodes might still be overwhelmed by the noise when using \autoref{eq:jag_depth}. \citeN{Fletcher2015c} and \citeN{Friedman2010} address this shortcoming, described below.

\citeN{Fletcher2015c} propose dynamically defining the maximum tree depth, with different depths possible in different parts of the same tree. By rephrasing the problem in terms of the support ``signal'' and the Laplace ``noise'', the authors define a minimum support threshold that ensures the signal-to-noise ratio \cite{VanDrongelen2006} is greater than one:
\begin{equation} \label{eq:dynamic_depth}
n' = \frac{\sqrt{2}|C|}{\epsilon} \enspace .
\end{equation}
If the estimated support of a node is below $n'$, the node is not split further. This prevents the tree from growing to a point where the noise added to each class count is likely to be larger than the count itself. \citeN{Fletcher2015c} estimate the support of a child node $i$ to be $\tilde{n_j}/b_j$, where $\tilde{n_j}$ is the estimated support of the parent node $j$ and $b_j$ is the branching factor of the attribute randomly selected by $j$ to split with. This allows each branch of the tree to continue growing (i.e. increase in depth) until the class counts are at risk of being overwhelmed by the Laplace noise added to them. Estimating the support of each node in this way is similar to how \citeN{Jagannathan2012} decide the maximum tree depth with $\log_b n-1$, except that it takes into account that different parts of the tree can have different numbers of branches. It also adapts to the size of $\epsilon$, changing the minimum support $n'$ based on how much noise is going to be added to each class count.

Both \citeN{Jagannathan2012} and \citeN{Fletcher2015c} use $b$ to estimate the support of nodes, making the assumption that the data will be roughly uniformly divided among all branches. This ensures that tree-building will terminate before the noise outweighs the class counts if the records are evenly distributed. If this assumption does not hold, and the records clump together in a small number of leaf nodes, this is still a good outcome if future unseen records follow the same distribution as the training records. The noise might completely overwhelm the class counts in small leaf nodes, but this will only affect a proportionally equally small number of future predictions.

\citeN{Friedman2010} solve the shortcoming of \citeN{Jagannathan2012}'s termination criteria (i.e., not taking into account the size of the noise) without using the branching factor $b$ at all. Instead, they directly query the data in the nodes as they build the tree. Recall that \citeN{Friedman2010} propose asking two queries in each node: one to return the support of the node, and the other to either return the best attribute to split the node with or to return the class counts if the node is a leaf. The latter query was addressed in \autoref{subsec:nonleaf} and \autoref{subsec:leaf}; the former query is used to enable Friedman and Schuster's third and final termination criteria. Interestingly, \autoref{eq:dynamic_depth} is the same formula that \citeN{Friedman2010} found through their experimentation; after approaching the problem of termination criteria from a different direction, \citeN{Fletcher2015c} discovered an explanation for why it worked. The main difference is whether $n'$ is compared to node $i$'s actual support $n_i$ or its estimated support $\tilde{n_i}$.

This termination criteria, along with \citeN{Rana2016}'s minimum support criteria discussed earlier, are the only data-dependent criteria proposed in the literature. Querying the support of each node allows the growth of different parts of the tree to be more precise than what \citeN{Fletcher2015c} could achieve. It comes at a cost however; part of the privacy budget $\beta$ must be spent on this query. It is debatable whether increasing how precise a termination criteria is is worth part of the budget; especially if it costs as much as half the budget, as seen in \autoref{eq:continuous} and \autoref{eq:rana_budget}. Fortunately the extra query that Friedman and Schuster ask in each node is not \emph{only} used for the termination criteria; they also use it when pruning the tree. We discuss this below in \autoref{subsec:pruning}.

\subsection{Pruning} \label{subsec:pruning}

After a tree has finished being built, pruning (described in \autoref{subsec:DTs}) can then be applied. In order to account for differential privacy and the modifications to a decision tree algorithm that that entails, conventional pruning strategies also need to be modified. \citeN{Friedman2010} propose one such modification for greedy decision trees, with some additional insight offered by \cite{Fletcher2015c}.

\citeN{Friedman2010} use the error-based pruning approach employed by C4.5 since it does not require additional data (data that could instead be used during the tree-building process, reducing the impact of the noise). Before performing pruning, they observe that due to the noise added to the support and class counts in each node, the sum of these counts in each level of the tree do not necessarily match the total size of the dataset. They therefore first ``normalize'' the support of each node in the tree so that the sum of all the nodes on each level of the tree match the number of records in the whole dataset. They then ``normalize'' the class counts in a similar fashion, making sure that the sum of the class counts in each node match the support of that node. The pruning is then applied as normal. The support $n_i$ of the nodes is normalized using the noisy counts gathered from the additional query the authors ask in each node (discussed in \autoref{subsec:nonleaf} and \autoref{subsec:termination}).

We take a moment to note that while this normalization process is helpful, it is possible that \citeN{Friedman2010} were motivated by flawed reasoning. They claim that the process of aggregating class counts in child nodes into the parent node during pruning causes the noise to ``add up'', leading to higher variance in the class counts. In fact, the noise is actually reduced, as described by the ``signal averaging'' phenomenon in signal-to-noise ratio theory \cite{VanDrongelen2006}. Taking a similar approach to how they approached termination criteria (see \autoref{subsec:termination}), \cite{Fletcher2015c} express the ratio between a parent node's support $n_i$ (i.e., the signal) and its total noise after summing the noise from its set of child nodes $J$ as:
\begin{equation} \label{eq:averaging}
\frac{n_i}{noise} = \epsilon \frac{\sum^J_j n_j}{|C|\sqrt{2|J|}} \enspace .
\end{equation}
Put simply, since the noise added to each count can add or subtract from the original count, summing multiple noisy counts together causes the noise to ``cancel out''. Thus the noise only scales with the square root of the number of child nodes, not linearly, nor super-linearly like is suggested by \citeN{Friedman2010}. This means that when pruning a differentially private tree, the class counts become less noisy (i.e., more reliable) as more pruning is performed. Of course the tree cannot be pruned too far, as the tree will start to lose discriminatory power and be less useful, just as is the case with conventional non-private trees.

When it comes to random decision trees, \cite{Fletcher2015c} observed an interesting phenomena: because nodes are split with a randomly selected attribute, it is entirely possible that the resulting child nodes are \emph{less} predictive than the parent node. In other words, a random split can actually make the tree worse. What this means is that pruning methods used by greedy trees no longer make sense for random trees. Instead the authors propose pruning in a way that targets the consequences of differential privacy: only pruning leaf nodes where the noise is greater than the class count ``signal''. Since the noise in a parent node is less than the sum of the noise in the child nodes (see \autoref{eq:averaging}), the number of prunes required is kept to a minimum, and the tree is kept as large as possible.

Continuing with the idea that a randomly chosen split might create child nodes with lower confidence than the parent node, \cite{Fletcher2015c} also propose altering the way that the labels of future records are predicted. For any given root-to-leaf path followed by an unlabeled record, the authors found that prediction accuracy rose if the node with the highest confidence is used to predict the record's label, regardless of where that node appears on the path.

\subsection{Multiple Trees} \label{subsec:forest}

Another design decision of any decision tree algorithm is the number of random decision trees to build. If the trees are built randomly, a large number of trees are required to result in a good classifier (i.e., a decision forest, see \autoref{subsec:random}). This is a core concept behind traditional, non-private random decision forests \cite{Fan2003,Geurts2006}, and is still true for differentially private random decision forests \cite{Jagannathan2012}. If the trees are built greedily, building more trees is not as necessary, but still beneficial \cite{Islam2011a}. This is especially true if they are built using bootstrapped samples and random subsets of the attributes, increasing the diversity of the trees and decreasing the model's susceptibility to small changes in the training data, which in turn reduces over-fitting \cite{Breiman2001}.

Unfortunately in the private scenario, each tree added to a forest comes with a cost; if each tree is built using the full dataset, the trees are not disjoint and cannot be queried in parallel (see \autoref{subsec:DP}). If the records are divided into disjoint subsets for each tree, the leaf nodes will have much lower support, and therefore less reliable majority class labels. The smaller a class count is, the larger the relative effect of adding $Lap(1/\epsilon)$ to it is. When using random decision trees, both strategies have been experimented with \cite{Jagannathan2012,Fletcher2015c,Fletcher2017}; in the case of greedy trees, only dividing the privacy budget has been tried \cite{Patil2014,Fletcher2015b}, with the feasibility of using disjoint data remaining an open question. We explore what conclusions can be made about building multiple differentially private decision trees below.

\citeN{Patil2014} were the first to take a differentially private greedy decision tree, such as the one proposed by \citeN{Friedman2010}, and expanding it into a forest. Their approach was to use bootstrapped samples in each differentially private tree, in the same way as Breiman's Random Forest algorithm \cite{Breiman2001}. In the eyes of differential privacy, sampling a dataset of size $n$ with replacement $n$ times (i.e. boostrapping) is almost identical to using the same data in every tree, and thus the privacy costs of the queries must be composed (\autoref{subsec:DP}). Some work has been done on calculating what privacy savings can be made by using bootstrapped data, since it is not \emph{exactly} the same as querying the original $n$ records \cite{Rana2016}, however thus far only approximations have been made, which do not guarantee differential privacy in the strongest sense. We explore these approximations later in \autoref{subsec:weak}.

In their empirical results, \citeN{Patil2014} demonstrated the infeasibility of their approach; even with a modest forest size of $\tau=20$ trees and a tree depth of $d=5$, the resulting classifier has poor prediction accuracy with very high variance. If part of the privacy budget $\beta$ is needed for one query in each level of a tree, for every tree, each query only receives
\begin{equation} \label{eq:greedyforest}
\epsilon = \frac{\beta}{d\tau}
\end{equation}
of the budget, leading to a very large amount of noise being added to each query for any reasonable budget $\beta$.

\citeN{Fletcher2015b} experimented further with making a forest of greedy trees, reducing the number of trees from \citeN{Patil2014}'s 20 trees to only four, and reducing the sensitivity of the queries to reduce the amount of noise required in each level of the trees. Their experiments indicated that a small number of trees often performed better than a single tree, despite the privacy budget having to be spread more thinly among multiple trees. When tested without the improved sensitivity however, the accuracy of the forest was basically the same as a single tree. These findings demonstrate that while increasing the number of greedy trees can be beneficial, it heavily relies on the noise in each query being reduced in other ways to compensate for dividing the $\beta$ among multiple trees.

Random decision trees, on the other hand, are only useful when part of a larger forest \cite{Fan2003,Geurts2006}. In the context of differential privacy, their greatest advantage over greedy trees is the complete removal of queries in the non-leaf levels of the trees. This allows the budget for each query to rise from that of \autoref{eq:greedyforest} to:
\begin{equation} \label{eq:randomforest}
\epsilon = \frac{\beta}{\tau} \enspace .
\end{equation}

\citeN{Jagannathan2012} used this advantage and searched for the ideal number of random decision trees to build differentially privately. They chose to use all the data in each tree, and divide the budget as shown in \autoref{eq:randomforest}. Experimentally, the authors concluded that $\tau=10$ was the ideal number of random trees to build. Unfortunately, there is no mathematical or heuristical explanation for this number; their only advice is that in their experiments, datasets with less than 500 records worked best with only $\tau=5$ trees, and that otherwise ten trees worked well. We suspect that this low number of trees was optimal because at $\tau>10$, the privacy budget was being divided among so many queries that the noise overwhelmed the signal.

Some theory was later developed by \citeN{Bojarski2015} that offered some probabilistic bounds on the prediction accuracy of a differentially private random forest, based on the number of trees and the depth of those trees. The authors released a preliminary version of their research on arXiv.org, presenting theorems for bounding both training error and testing error (in other words, classifying training data and non-training data). Notably, they propose that to correctly classify most new (non-training) data, a logarithmic number of random decision trees $\tau$ is needed, relative to the number of training records $n$. In their experiments, the number of trees went as high as 21, but also as low as one. This lines up reasonably well with \citeN{Jagannathan2012}'s recommendation of using $\tau=10$. \citeN{Rana2016} implemented the findings of \citeN{Bojarski2015} as a comparative decision forest in their experiments, with mixed results. Since proofs of the proposed bounds have not yet been released, it is difficult to discern their efficacy. 

Taking a heuristical approach, \citeN{Fletcher2015c} proposed basing the number of random trees on two factors: the number of attributes $m$; and the ``minimally acceptable tree size'', defined as a tree with a depth of $d=3$. The authors developed a heuristic whereby the the budget assigned to each tree $\epsilon=\beta/\tau$ would not be so small that their minimum support threshold $n'$ (see \autoref{eq:dynamic_depth}) would prevent a tree from splitting at least twice; in other words if 
\[
\frac{|C|\sqrt{2}}{\epsilon} > \frac{n}{b^2} \enspace , 
\]
where $b$ is the average branching factor. What this means is that the number of trees $\tau$ has an upper bound, so that $\epsilon$ can be large enough to allow for minimally acceptable tree sizes.

They additionally maximized tree diversity by requiring that each tree's root node use a unique attribute to split with, with another upper bound on the number of trees being the number of attributes; $\tau\leq m$. Any additional trees would contribute less unique information on average than the first $m$ trees, causing inefficient spending of $\beta$. A heuristical solution to Jagannathan et al.'s difficulties in defining the number of trees was therefore to define it as:
\[
\tau = \min\left( \frac{\beta n}{\sqrt{2}|C|b^2},~m \right) \enspace .
\]

\citeN{Fletcher2017} researched building an ensemble of differentially private random trees from a different angle: they proposed using sampling without replacement (i.e., disjoint subsets of dataset $x$) in each tree, rather than using all of the data in each tree. This strategy has some of the benefits of the bootstrap sampling originally proposed by \citeN{Breiman2001} (i.e., sampling \emph{with} replacement) in that it reduces the variance caused by individual records, but the authors' main reason for sampling without replacement is to allow the algorithm to use parallel composition (\autoref{def:parallel}). This is interesting, because in many cases (such as counting queries) the difference between using small subsets of data in parallel versus using small fractions of the privacy budget with all the data is a zero-sum game. You can add $Lap(k/\epsilon)$ noise to a count of size $n$, or you can add $Lap(1/\epsilon)$ noise to a count of size $n/k$; the noise remains proportional to the data. What \citeN{Fletcher2017} found however, is that this did not hold true when using the Exponential mechanism with smooth sensitivity (see \autoref{subsec:smooth}), since a larger budget $\epsilon$ improved both the numerator and denominator of \autoref{eq:expo}, while a larger $j$ (which scales asymptotically linearly with data size $n$) only improved the denominator.

Another natural consequence of reducing the noise in the leaf nodes with smooth sensitivity is that it means a smaller proportion of the budget is required to get the same level of accuracy. \citeN{Fletcher2017} found that this allowed their algorithm to build a much larger forest, gaining the benefit of 100 votes\footnote{By ``votes'', we mean the predictions made by each individual tree in the forest, with the most popular prediction overall being the final prediction.} per predicted class label instead of the 10 offered by \citeN{Jagannathan2012}.

\subsection{Weakening the Privacy Requirements} \label{subsec:weak}

Jagannathan and Pillaipakkamnatt, this time joined by Monteleoni, extended the work they did with Wright the previous year to make a semi-supervised classifier in a \emph{semi-private} scenario \cite{Jagannathan2013}. The scenario they describe is one where the user has a large number of unlabeled non-private (i.e., public) records, and a small number of labeled private (i.e., sensitive) records. The authors propose an algorithm that uses the unlabeled records in two ways. First, to find dense regions of the data hypercube and partition those regions when building decision trees, aiming to spread the records evenly among the leaf nodes. Second, to use a small differentially private random forest built from the private (labeled) data to classify the non-private (unlabeled) data, and then build a large non-private decision forest with the non-private data. The final classifier is then the union of the two decision forests.

The main weakness of this approach is its applicability in the real world. We question how often a user would find themselves in a situation that meets all of the criteria: a large number of records that are both unlabeled and non-private, and a small number of records that are both labeled and private, with both types of records being from the same distribution.

Taking a different approach, \citeN{Rana2016} proposed a greedy decision forest algorithm that used a weaker definition of differential privacy. By weakening the requirements of differential privacy, it allowed the forest to be much larger than previous forests made of greedy decision trees \cite{Patil2014,Fletcher2015b}. Rather than hiding whether an individual's record is in a dataset with high probability, Rana et al.'s weakened definition of differential privacy makes sure that any estimate a user makes about a specific attribute value has the same variance as it did before the user made any queries, with high probability. In other words, their definition prevents an individual's values being discovered (attribute linkage), but not whether or not the individual is in the dataset (table linkage) \cite{Fung2010}. In the paper, the scope of this definition is limited to an ensemble of decision trees using bootstrapped samples; no work has yet been done to extend it to other scenarios.

Similar to \citeN{Patil2014}, the authors propose a greedy decision forest algorithm that builds $\tau$ trees in the same fashion as \citeN{Friedman2010}, where each tree uses a random sub-sample of the total dataset. Rather than using a budget of $\epsilon=\beta/\tau$ in each tree, Rana et al. distinguish their algorithm from others by allowing the user to manually define $\epsilon$ (where $\epsilon\leq\beta$), with each tree using that amount. They demonstrate that their weakened definition of differential privacy is preserved by the resulting decision forest as long as the number of trees is limited to:
\begin{equation}\label{eq:rana}
\tau \leq \frac{1}{p(1-p)} \left(n'(n'+1)+\frac{1}{\epsilon^2}\right) \enspace ,
\end{equation}
where $n'$ is the minimum support of leaf nodes, $p$ is the probability of any randomly chosen class label being the least common class label, and $\epsilon$ is the user-defined privacy budget per tree. As discussed in \autoref{subsec:termination}, $n'$ is user-defined here. Note that Rana et al. only present formulas for scenarios with a binary class.\footnote{They also provide several additional parameters in the paper, which we generalize here to avoid minutiae.} \autoref{eq:rana} only bounds the forest size in terms of preserving the class attribute; Rana et al. also define an upper bound on the number of trees to ensure no leakage of \emph{non-class} attribute values, however in practice this bound is much higher than the bound based on the class attribute so we do not include it here.

\subsection{Query Efficiency} \label{subsec:efficiency}

In this section, we use \citeN{Friedman2010}'s differentially private greedy decision tree as a case study in how to efficiently query a dataset when designing a data mining algorithm, and how to save parts of the privacy budget where possible.

In \autoref{subsec:nonleaf}, we discussed how \citeN{Friedman2010} asked two queries per node. The authors divided the privacy budget $\beta$ evenly among all the queries, with each query having $\epsilon=\beta/2d$, where $d$ is the depth of the tree. One natural question might be, ``is the first query in each node (which outputs the support) useful enough to be worth half the budget?''. While no research has directly explored this question, a differentially private decision tree proposed by \citeN{Mohammed2015} takes a very similar approach to Friedman and Schuster, but removes the first query in each node. This allows the splitting function in each node to have twice as much of the budget, $\epsilon=\beta/d$. Mohammed et al. demonstrated that not only can a differentially private greedy tree be built with only one query per node, but that it actually achieves higher prediction accuracy than Friedman and Schuster's version that has the more precise termination criteria and pruning.

While ultimately resulting in lower prediction accuracy, Friedman and Schuster demonstrated an important concept when designing differentially private algorithms: using queries that can achieve multiple purposes at once. By using the support of each node in two separate components of their algorithm (termination criteria and pruning), the authors got more ``bang for their buck''. Differential privacy is immune to post-processing \cite{Dwork2014}, so outputs that can be used multiple times to improve a classifier are essentially ``free'' optimizations.

We can also observe that in \autoref{subsec:termination}, Friedman and Schuster's \emph{Criteria 3b} will never be used by the algorithm if $n/b^d$ is larger than the standard deviation of the Laplace noise, $\sqrt{2}/\epsilon$; one of the other two criteria will always occur first. If this is the case for a particular dataset, querying the support of each node is no longer useful for the termination criteria, and is only useful for pruning. The query was of questionable utility when it had two purposes (termination and pruning), let alone one, and should probably be skipped in this case, saving half the budget. We can take this idea further and ask ourselves: would we even want to terminate tree-building at $d\leq2$? If not, then why not only query the support of nodes when $d>2$, thus reducing the number of queries by two? 

We should also first check the two data-independent criteria (\emph{Criteria 1} and \emph{Criteria 2}) at each level, and maybe not query the support if one of the other criteria is already terminating the tree's growth. On the topic of the other two criteria, note how terminating because all the attributes $m$ have been tested is unlikely to occur unless $m$ is less than the user-defined maximum depth $d$. However, this termination criteria differs from the minimum support criteria in that it does not \emph{cost} us anything, since it does not rely on the data. We can afford to have redundant components in the algorithm, as long as they do not cost us.\footnote{Of course, computational complexity may also want to be considered, however in privacy scenarios we often want to maximize the utility-to-privacy trade-off, and are unwilling to compromise for the sake of other trade-offs like utility-to-complexity.}

If the user still wishes to query the support of each node (or perhaps ask some other non-compulsory query), they can consider whether every query should have an even proportion of the total budget. To the best of our knowledge, all research around differentially private trees has divided $\beta$ evenly, such as $\epsilon=\beta/2d$ or $\epsilon=\beta/\tau$. This is not required however, and it is worth considering if it is more important that certain queries are less noisy, and that other queries can afford to be noisy without sacrificing too much utility.

These sorts of heuristics can lead to large budget savings, and thus a better classifier. While we used \citeN{Friedman2010}'s algorithm as a case study, similar ideas can be used when designing any differentially private algorithm.

\section{Bringing it all Together: Implementations} \label{sec:together}

\autoref{tab:properties} summarizes the main properties of all the algorithms discussed. While no table could properly capture all the details of these algorithms, \autoref{tab:properties} provides a good overview and can act as a quick reference. \autoref{fig:jag_vs_fri} and \autoref{fig:smooth_vs_jag} provide some context for the ``Low/Medium/High'' categories in the ``Prediction Accuracy'' column.

\begin{table}
\renewcommand{\arraystretch}{1.4}
\tbl{A comparison of the main properties of the differentially private decision tree algorithms.}{
\centering
\footnotesize
\begin{threeparttable}
\begin{tabular}
	{>{\centering\arraybackslash}m{2.2cm} >{\centering\arraybackslash}m{1.2cm} >{\centering\arraybackslash}m{1.2cm} >{\centering\arraybackslash}m{1.6cm} >{\centering\arraybackslash}m{1.0cm} >{\centering\arraybackslash}m{1.5cm} >{\centering\arraybackslash}m{1.5cm}}
	\noalign{\smallskip}\hline\noalign{\smallskip}	
	\textbf{Classifier} & \textbf{Budget per Query} & \textbf{Tree Type} & \textbf{Tree Depth $d$} & \textbf{Forest Size $\tau$} & \textbf{Handles Cont. Attributes} & \textbf{Prediction Accuracy\tnote{4}} \\
	\noalign{\smallskip}\hline\noalign{\smallskip}
	Blum, Dwork, McSherry \& Nissim [2005] \tnote{1} & $\frac{\beta}{2md-1}$ & Greedy & $m$ & 1 & No & Very Low\tnote{5} \\
	\noalign{\smallskip}
	Friedman \& Schuster [2010] \tnote{2} & $\frac{\beta}{(2+n)d}$ & Greedy & 5 & 1 & Poorly & Low \\ 
	\noalign{\smallskip}
	Mohammed, Barouti, Alhadidi \& Chen [2015] \tnote{2} & $\beta/d$ & Greedy & 4 & 1 & No & Medium\tnote{5} \\
	\noalign{\smallskip}
	Patil \& Singh [2014] \tnote{2} & $\frac{\beta}{\tau(2d-1)}$ & Greedy & 5 & 20 & No & Very Low \\
	\noalign{\smallskip}
	Fletcher \& Islam [2015a] \tnote{3} & $\frac{\beta}{\tau(2d-1)}$ & Greedy & 5 & 4 & No & Low \\
	\noalign{\smallskip}
	Rana, Gupta \& Venkatesh [2016] \tnote{2} & $\beta/2d$ or $\beta/3d$ & Greedy & Unspecified & $500+$ & Yes & High\tnote{6} \\
	\noalign{\smallskip}
	Jagannathan, {Pillaipakkamnatt} \& Wright [2012] \tnote{2} & $\beta/\tau$ & Random & Maximized Diversity (Discrete) & 10 & Yes & Medium \\
	\noalign{\smallskip}
	Jagannathan, Monteleoni \& {Pillaipakkamnatt} [2013]\tnote{2} & $\beta/\tau$ & Random & Maximized Diversity (Discrete) & 5 then 200 & Yes & High \\
	\noalign{\smallskip}
	Fletcher \& Islam [2015b] \tnote{3} & $\beta/\tau$ & Random & Dynamic & $\leq m$ & No & Medium \\
	\noalign{\smallskip}
	Fletcher \& Islam [2017] \tnote{3} & $\beta$ & Random & Maximized Diversity & 100 & Yes & High \\
	\noalign{\smallskip}\hline
\end{tabular}
\begin{tablenotes}
		\item[1] Code available at \emph{https://www.microsoft.com/en-us/research/project/privacy-integrated-queries-pinq/} .
		\item[2] No code has been made publicly available.
		\item[3] Code available at \emph{https://samfletcher.work/code/} .
		\item[4] Prediction accuracy is categorized into quartiles of average relative empirical performance, calculated using the prediction accuracy results presented by each of the authors across a variety of datasets and $\beta$ values.
		\item[5] Our best estimation; the authors did not empirically test the algorithm.
    \item[6] With a weaker definition of differential privacy.
\end{tablenotes}
\end{threeparttable}
}
\label{tab:properties}
\end{table}

In some cases, such as some of the cells in the ``Tree Depth $d$'' and ``Forest Size $\tau$'' columns, the values provided are not inherently part of the proposed algorithm, but are instead the recommended values for parameters that are user-defined. For example, \citeN{Mohammed2015} recommend setting the user-defined depth to $d=4$, but since they only propose a decision tree algorithm and not a forest, their forest size is inherently $\tau=1$. In the case of \citeN{Rana2016}'s algorithm, the authors do not specify $d$ in their paper, and the size of their forest is defined by \autoref{eq:rana} but is usually $\tau>500$. 

Some cells are labeled as ``Maximized Diversity'', by which we refer to the combinatorial reasoning put forth by \citeN{Fan2003} to maximize the attribute diversity of the trees, discussed in \autoref{subsec:termination}. The inclusion of ``(Discrete)'' in these cells indicates that the authors only accounted for discrete attributes, and not continuous attributes. For \citeN{Jagannathan2013}'s work, we describe the forest size as ``5 then 200'', referring to the fact that the algorithm builds two separate decision forests; a small differentially private one, and then a large non-private one. The algorithm proposed by \cite{Fletcher2015c} uses a dynamic tree depth, with the minimum support threshold of each branch being updated as the tree grows. The ``Handles Cont. Attributes'' column lists whether an algorithm handles continuous attributes, or only discrete attributes.

For all algorithms in \autoref{tab:properties}, the mechanisms used to preserve privacy do not affect the computational complexity of the algorithm. The Laplace mechanism simply adds noise to outputs, and the Exponential mechanism probabilistically selects an output from a set of outputs that non-private versions of the algorithms would need to calculate anyway. Interestingly, because these algorithms focus on querying the data as few times as possible, the computational complexity indirectly benefits from design decisions made in the interest of privacy. In each case, the computational complexity is equal to the complexity of the same algorithm with the privacy-preserving components removed. Assuming $n,m,\tau \gg d$, the computationally complexity of the greedy algorithms in \autoref{tab:properties} is $O(nm^2\tau)$, $O(n)$ for \citeN{Fletcher2017}'s random forest, and $O(n\tau)$ for the other random algorithms.

While the error bounds of individual queries is trivial to calculate -- it is described by the same Laplace probability density function used to add the noise -- little research has been done on evaluating the error bounds of more complex combinations of multiple queries. \citeN{Hay2016} proposed a method for empirically measuring how much error is incurred, rather than relying on theoretical bounds, however the method has not yet been extended for combinations of queries.

The details of all the algorithms summarized in \autoref{tab:properties} are discussed and compared throughout \autoref{sec:main}.

\subsection{Choosing the Privacy Budget} \label{sec:budget}

In all of the algorithms talked about in this survey, the privacy parameter $\beta$, and how it is subdivided into $\epsilon$ values for each query, determines how much noise is added to the output of queries. This extends to all applications of differential privacy; $\beta$ and $\epsilon$ are at the core of its definition. Despite this, very little work has been done in how to choose an appropriate value for $\beta$ in the real world, both for ensuring comparable utility between private and non-private queries, and for providing an appropriate amount of privacy protection to each of the individuals in the data. How large $\beta$ can be is a ultimately a social problem that depends on how highly the individuals in a dataset value their privacy. Dwork et al. themselves have left this question up to the discretion of the data curator and their particular circumstances \cite{Dwork2014}. To the best of our knowledge, two papers have provided practical guidelines for choosing $\beta$; one from the perspective of utility \cite{Vu2009} and one from the perspective of privacy \cite{Hsu2014}.

Using the context of clinical trials, \citeN{Vu2009} provided several results in 2009 for how to use hypothesis testing in a scenario where sensitive information needs protecting. They focus on the sample size of the test, and demonstrate how many more records are needed to produce the same quality of results that the test would produce if privacy was not a concern. The authors do so in terms of two main factors: the confidence level of the hypothesis test (type I error) and the power of the hypothesis test (type II error). The equations for this scaling of the number of records, dubbed the ``sample size correction factor'', can be found in \citeN{Vu2009}. If a user is able to estimate how many samples they would need to achieve strong results for their hypothesis test, they can use Vu and Slavkovic's work to achieve the same results, but differentially privately.

Rather than focusing on the sample size required to negate the influence of the added noise, \citeN{Hay2016} proposed a method for empirically measuring how much noise is added to any differentially private results. Instead of relying on the theoretical bounds outlined by the mathematics of differential privacy, \citeN{Hay2016} enable a user to measure the actual noise added in practice. Unfortunately the method can only be applied to simple queries independently -- further research is required to extend it to sequences of queries, or larger structures made up of many ``building block'' queries, such as machine learning models.

On the other side of the utility-privacy trade-off, \citeN{Hsu2014} proposed the first practical methodology for deciding how small $\beta$ needs to be to appropriately protect people's privacy for a given scenario. They do so by modeling the problem in terms of money; the monetary budget of the user collecting or using the sensitive data. In any given scenario, each individual included in the data is at a certain amount of risk. By estimating the probability of an individual's privacy being maliciously breached, and the monetary cost associated with that breach, the user can calculate the average risk per individual in the data. Since the probability of a breach is directly tied to $\epsilon$ (i.e., the presence of a record can change the output with probability proportional to $\exp(\epsilon)$, see \autoref{def:DP}), the user can balance how many samples they collect/use, how large $\epsilon$ (and therefore $\beta$) can realistically be, and how much money they are willing to spend (either on paying people to participate in the study or on compensating the individuals whose privacy is breached). Such a model offers practical guidelines for how large or small the privacy budget can realistically be.

\section{Private Data Publishing with Decision Trees} \label{sec:publish}

When releasing a differentially private model, the data curator is releasing a particular interpretation of the original data, that enables certain kinds of inference, but prevents other kinds. For example, a tree has one outcome variable that it is predicting, and does not enable a user to use that tree to predict a different variable. This type of data release is an example of ``privacy-preserving data mining'' \cite{Aggarwal2008c}, where data mining is applied directly to the original data using a private mechanism, and the resulting private model is released to the public. Sometimes the data curator would prefer to release a private version of the \emph{data itself} to the public. This scenario is an example of ``privacy-preserving data publishing'' \cite{Fung2010}, where the data is kept as similar as possible to the original data \cite{Fletcher2014} while still protecting privacy. While this survey focuses on the construction of differentially private decision trees, we take a moment to briefly discuss the other side of the coin: how traditional, non-private decision trees can be used to publish private data. This includes privacy-preservation strategies other than differential privacy as well.

While differential privacy is becoming the de-facto standard for modern privacy preservation \cite{Clifton2013}, other forms of privacy preservation exist. The main competing approach is $k$-anonymity \cite{Sweeney2002}, which uses the idea of ``generalization''. In a privacy context the term ``generalization'' refers to making attribute values more ``general''; that is, less granular.\footnote{Not to be confused with the definition of ``generalization'' that describes a classifier's ability to generalize to unseen data.} A classic example is to generalize birth dates to birth years, or to generalize street addresses to suburbs. $k$-anonymity states that every record in a dataset should have all the same values as at least $k-1$ other records in the dataset. Fung et al. proposed a practical implementation named ``Top-Down Specialization'' \cite{Fung2005,Fung2007} that achieves $k$-anonymity by first making every value in the dataset as general as possible, and then building a greedy decision tree that iteratively increases the granularity of attributes. Attributes were selected to be made more granular (i.e., specialized) if they had high information gain compared to the amount of privacy they leaked. In other words, $k=N$ at the root node, and then $k$ is reduced with each split as efficiently as possible, terminating when $k$ reached the user-defined minimum. After the decision tree was built, a synthetic $k$-anonymous dataset could be released that used the specializations chosen by the tree.

Interestingly, Top-Down Specialization was re-purposed in 2011 \cite{Mohammed2011} to provide differential privacy instead of $k$-anonymity. Instead of specializing attributes based on their utility-privacy trade-off (in the node splitting function), only utility is considered. The resulting tree is then used to create segments of generalized records, whose frequency can be differentially privately outputted. Thus a frequency table of generalized records can be released. The reasoning for this approach was that generalization offers something that adding Laplace noise does not; values in the dataset cannot be outright incorrect, merely less granular. In scenarios where misleading information must be avoided at all costs, this approach can be very appropriate. When building a non-private decision tree from the generalized data, the prediction accuracy was similar to the accuracy achieved by Friedman and Schuster's algorithm \cite{Friedman2010}.

Another type of privacy preservation is noise addition, in which noise is selectively added to attribute values to hide sensitive information \cite{Agrawal2000}. Several privacy-preserving decision tree algorithms have been proposed that add noise in such a way that the original patterns found by the decision trees are preserved in the anonymized data \cite{Islam2011,Fletcher2015a}. Noise addition offers less rigorous protection of people's privacy than $k$-anonymity however, and neither technique offers mathematical guarantees like differential privacy. Perhaps most egregiously, they provide no protections against malicious users who have supplementary information (information from other sources about the people in the dataset). Without clearly definable guarantees for each individual, it is difficult for noise addition or $k$-anonymity to be implemented in a way that meets the real-world privacy policy requirements of governments and businesses. A detailed comparison of the advantages and disadvantages between $k$-anonymity and differential privacy was performed by Clifton and Tassa in 2013 \cite{Clifton2013}.

\citeN{Mivule2012} also proposed a framework that iteratively checks the quality of a differentially private dataset with an AdaBoost decision forest \cite{Freund1999}, updating the dataset until the forest achieves acceptable prediction accuracy. Their framework lacked detail however, and their results were mixed.

As differentially private algorithms become more specialized, different algorithms will perform better in different circumstances, depending on characteristics of the underlying data and the problem domain. \citeN{Kotsogiannis2017} propose using a decision tree to deduce the most appropriate differential privacy algorithm to use for any given scenario, out of a collection of possible algorithms. This ``meta-tree'' facilitates the creation and application of specialized algorithms, rather than limiting users to more general-use algorithms. Other research has discovered that in some domains it can be very difficult to achieve acceptable quantitative results with differential privacy, and that extracting qualitative information may be more beneficial \cite{Bai2018,Hu2015}.

Many other techniques have been proposed for publishing differentially private data, though not with decision trees and therefore outside the scope of this work. They often involve partitioning the original dataset in a way that balances utility with privacy, then creating aggregated records from noisy histograms of the partitions \cite{Blum2013,Li2014}. The utility of these techniques suffers exponentially when the number of attributes increases beyond four, making them unusable in most real-world scenarios (barring an equally exponential increase in the number of records). PrivBayes \cite{Zhang2014} focuses on avoiding this curse of dimensionality, by using a Bayesian network to publish synthetic data from noisy marginals. We refer the reader to \citeN{Ji2014}'s and \citeN{Zhu2017a}'s surveys for more details.

\section{Looking Forward} \label{sec:future}

Looking forward into the future of differentially private trees, several open questions catch our attention. While random trees currently reign supreme when it comes to prediction accuracy (see \autoref{fig:jag_vs_fri}), the recent developments with smooth sensitivity \cite{Fletcher2017} offer some promising possibilities for greedy trees. We wonder if using smooth sensitivity would lower the noise enough to allow the Exponential mechanism to output the best splitting attribute (see \autoref{subsec:nonleaf}) more reliably at low $\epsilon$ values. Not only would this improve single greedy trees, but it might open the way for ensembles of greedy trees to perform well; currently 20 or even less greedy trees divides the privacy budget too much to be useful \cite{Fletcher2015b,Patil2014}.

Greedy decision trees may also be able to be improved by reducing the number of queries required. For example, algorithms that do not require pruning have been proposed in non-private scenarios \cite{Hothorn2006}, that contain strategies that could be applicable in differentially private tree algorithms. Extending these algorithms to handle continuous outcome variables would be another interesting direction to explore, enabling the creation of differentially private regression trees.

The optimal depth of differentially private greedy trees is another open question. The depth of their random counterparts has been explored much more thoroughly, as we can see in \autoref{subsec:termination}. This question becomes even more interesting when considering the cost of the queries in each non-leaf level of the tree; the deeper the tree grows, the more the budget must be divided. This of course presupposes that there is not a creative way of using disjoint data in each level of the tree; such a discovery would open up a wealth of new possibilities.

\autoref{subsec:forest} demonstrates that there is some amount of disagreement about how many trees are needed in a differentially private random forest. \citeN{Jagannathan2012} claim that $\tau=10$ trees is enough to achieve high accuracy, citing the findings of \citeN{Fan2003} as evidence. Conversely \citeN{Fletcher2017} found that on the order of 100 trees was preferable, demonstrating their claim empirically. This matches more closely with the work of \citeN{Geurts2006}, who used 100 to 1000 random trees. More research in this area may elicit some heuristics for defining $\tau$, or perhaps more rigorous mathematics for finding the optimal $\tau$.

\citeN{Rana2016} explored weakening the definition of differential privacy, which led to greatly improved results (discussed in \autoref{subsec:weak}). They proposed one strategy for how to weaken the definition, but there is no reason to suspect that there are not other possible definitions. The original privacy definition proposed by \citeN{Dwork2006} has been thoroughly investigated and is mathematically rigorous (see \autoref{subsec:DP}), and this high standard should be required of any competing definition. \citeN{Rana2016}'s definition (and any future definitions) would thus require more research before they can be considered true alternatives to the original definition when the privacy of real people is at stake. If such an alternative was developed though, the improvements in data mining quality could be substantial.

In \autoref{subsec:efficiency} we discussed how small savings can be made with the privacy budget by ``cutting corners'' where possible. While an elegant algorithm might require $g$ queries, heuristics can allow us to remove $h$ unnecessary queries by either inferring the results from the other $g-h$ queries, or relying on data-independent estimates where accurate results are not absolutely necessary. Increasing the budget per query from $\epsilon=\beta/g$ to $\epsilon=\beta/(g-h)$ is likely worth any increases in algorithmic complexity caused by adding additional steps or heuristics. As differentially private decision trees are further honed in the future, small improvements such as these will be key to optimizing the results in the real world.

The final open area that catches our attention is the idea of spending unequal quantities of the privacy budget on different queries. So far in the literature, all algorithms have assumed that all $g$ queries will use $\beta/g$ of the budget, implicitly assuming that all the queries are of equal importance to the success of the algorithm, or require equal levels of accuracy to be useful. We suspect that some queries, however, can achieve their purpose in the algorithm with more added noise than other more ``fragile'' queries. This would be an interesting idea to explore in the future.

\section{Conclusion} \label{sec:conclusion}

When sensitive data -- data that risks the privacy of the people it describes -- needs to be data mined, decision trees lend themselves well to the strict requirements of differential privacy. We have surveyed the current literature of differentially private decision trees, breaking down the state-of-the-art algorithms into their constituent parts. Termination criteria, pruning strategies, building an ensemble of trees, greedy heuristics in non-leaf nodes, and querying the labels in leaf nodes are some of the factors that must be considered when making a high-performing decision tree algorithm. We explored each of these factors in turn, as well as others. We compared and contrasted the design decisions made by the authors, evaluating their effectiveness at discovering predictive knowledge from sensitive data.

One theme that re-occurred throughout our discussion was the ever-present conflict between privacy and utility. Whenever the sensitive data is queried, there is a cost associated, and that cost needs to be weighed against the benefits. Those benefits depend on the aim of the algorithm; is it aiming to discover knowledge, build a model with high prediction accuracy, both, or something else? These aims each lend themselves to slightly different strategies, changing the cost-benefit analysis that the user makes when deciding when their algorithm should query the data. Depending on how a query is asked, its sensitivity can also change drastically. As we saw in \autoref{subsec:smooth}, even changing a query slightly from ``what are the class counts in this leaf node?'' to ``what is the majority class label in this leaf node?'' allows for a much less noisy answer to be outputted.

Balancing the requirements of differential privacy with a decision tree algorithm that optimizes utility is a delicate process, and one that should be done with great care. It is our hope that the reader has gained a thorough understanding of the nuances involved in designing a differentially private decision tree algorithm, and we eagerly await the progress that the future holds.

\bibliographystyle{ACM-Reference-Format-Journals}
\bibliography{My_Collection}

\received{November 2016}{X 2018}{Y 2019}

%
%

\end{document}